\documentstyle[epsfig,11pt]{article}
\begin{document}

\vspace*{-1.8cm}
\begin{flushright}
{\bf LAL 04-114}\\
\vspace*{0.1cm}
{October 2004}
\end{flushright}
\vspace*{0.8cm}

\begin{center}
{\LARGE\bf Alternative IR geometries for TESLA\\ with a 
small crossing angle\footnote{Contributed to the 
machine-detector interface session of LCWS04.}}
\vspace*{0.5cm} 

{\large\bf R.~Appleby\footnote{r.b.appleby@dl.ac.uk}, D.~Angal-Kalinin}\\
{\it ASTeC, Daresbury Laboratory, Warrington, WA4 4AD, England}
\vspace*{0.2cm}

{\large\bf P.~Bambade, B.~Mouton}\\
{\it LAL, IN2P3-CNRS et Universit\'e de Paris-Sud, B\^at. 200, BP 34, 91898 Orsay Cedex, France}
\vspace*{0.2cm}

{\large\bf O.~Napoly, J.~Payet}\\
{\it DAPNIA-SACM, CEA Saclay, 91191 Gif/Yvette Cedex, France}
\vspace*{0.5cm} 
\end{center}

\begin{abstract}
The formulation of hybrid crossing angle schemes has been a recent development
of the TESLA collision geometry debate. Here we report on two such schemes, characterised by either a small vertical or horizontal beam crossing angle.
\end{abstract}

\section{Introduction}

The specification of the International Linear Collider (ILC) states
the need for two separate interaction regions (IRs) with comparable
performances and physics potential for e$^+$e$^-$-collisions. The time
structure of the bunch trains does not require colliding the beams 
with a crossing angle in the case of the superconductive technology, which has 
now been chosen for the accelerator, and there is hence a flexibility in the design. 
It is believed that both IRs with and without a crossing angle can lead to acceptable
designs. There are however specific advantages and disadvantages in each, for the machine  
as well as for some aspects of the physics potential~\cite{xanglepaper}. 
Moreover, if a crossing angle is used its magnitude is an important parameter to 
optimise. An additional consideration is the requirement to enable~$\gamma \gamma$ 
collisions as an option at one IR in the future. A large (still to be defined) 
crossing angle will most certainly be needed for the corresponding IR, while making sure not
to compromise its e$^+$e$^-$ capabilities. A balanced scenario which could be
attractive in this context would be to use a smaller or even null crossing angle at the other IR. In the TESLA technical design report~\cite{teslatdr} a head-on collision geometry was actually specified, but the extraction of the beamstrahlung photon flux and spent beam was found to be problematic during the evaluation conducted in 2003 by the ILC TRC, which even highlighted it as a level 2 ranking item in its final report~\cite{ILCTRC}.

In this paper, we describe two new so-called hybrid schemes featuring
small~$\mathcal{O}$(10$^{-3}$ rad) crossing angles. They are attempts to
maintain the advantages of the head-on geometry while resolving some of its
weaknesses. The first scheme, which uses a small vertical crossing angle, was 
originally suggested by Brinkmann~\cite{brinkmanntalk} to reduce the power deposition 
from beamstrahlung on the septum blade used in the extraction of the spent beam.  
In the solution presented in section~\ref{vert}, this vertical crossing angle is 
combined with modified optics in the final focus to improve the chromatic properties 
in the transport of the low energy tail of the spent beam and thereby reduce losses 
in the extraction channel. The second scheme, which uses a horizontal crossing angle, 
was first developed in the context of CLIC~\cite{napoly}. The solution presented in 
section~\ref{hori} is an adaptation to the TESLA project.

Both schemes are discussed is this paper in the spirit of initial proofs of principle. Investigations were only carried out at a centre-of-mass energy of 500 GeV and $L^*$, the distance between the last quadrupole and the interaction point, was taken to be 4.1m. Both geometries need 
further work and development. This will be pursued in the coming months to fully assess 
their feasibility for the IR which will not later be upgraded to $\gamma \gamma$
collisions, for centre-of-mass energies up to 1 TeV.

\section{The small vertical crossing angle scheme}
\label{vert}

The TESLA extraction scheme with a head-on collision described in the TDR~\cite{teslatdr} suffers from the problems of septum irradiation~\cite{septumrad} and the loss of low energy tail particles. The total power radiated on the septum blade was found to be unacceptable~\cite{septumrad}, and higher than the estimate in the TESLA TDR, when calculated for a realistic beam using start-to-end simulations. Analysis of the transport of the post-IP beam down the extraction line also revealed that the loss of charged particles can reach unacceptable levels in the septum blade region. This loss is a consequence of a beam size increase resulting from overfocusing of the low energy disrupted beam tail by the strong final doublet. The solution to avoid these problems is 
twofold~\cite{brinkmanntalk}. 

The septum irradiation problem is solved by introducing a small vertical crossing angle to shine the beamstrahlung away from the septum blade. The required vertical crossing angle can be estimated from the vertical photon distribution~\cite{septumrad} and the upper limit considered reasonable for the power deposition. An angle of $\sim$0.3mrad should be sufficient. 

\begin{figure}[h]
\begin{center}
\begin{minipage}{0.48\textwidth}
\includegraphics[width=7.7cm]{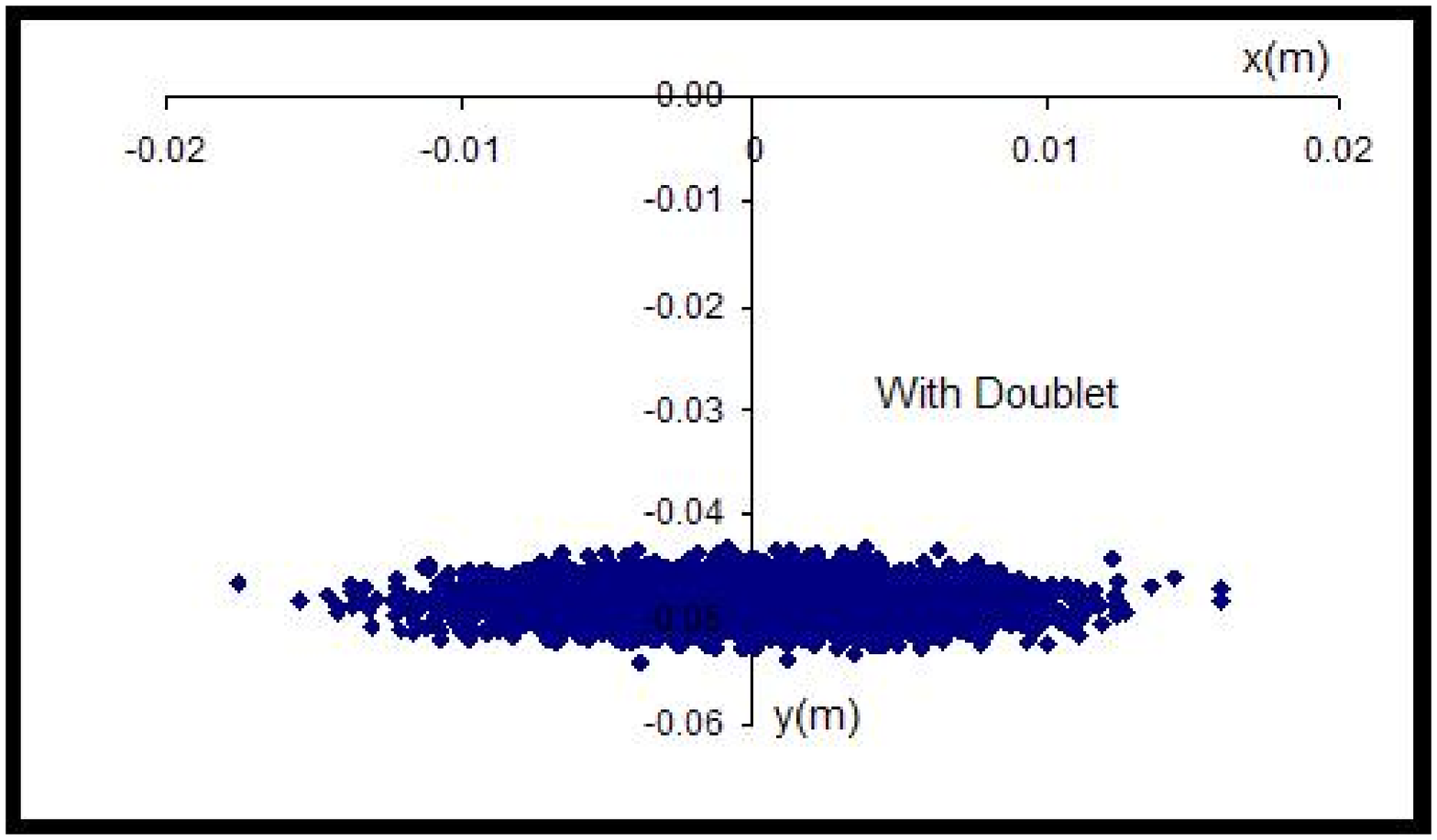}
\end{minipage}
\hfill
\begin{minipage}{0.48\textwidth}
\includegraphics[width=7.7cm]{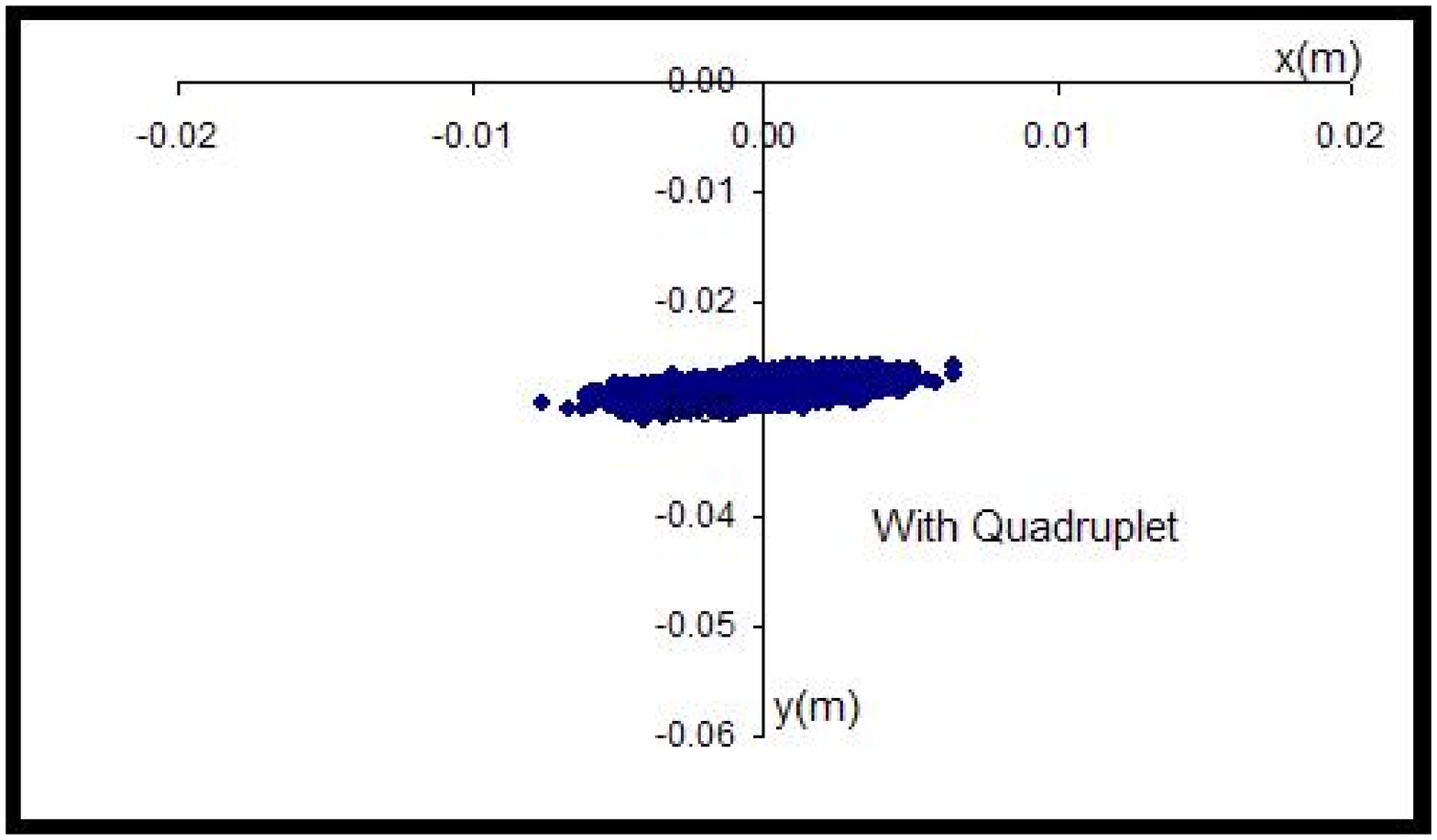}
\end{minipage}
\caption{\it The disrupted beam size in the region of the extraction line 
septum magnet. The right(left) plot shows the beam size obtained using the doublet(quadruplet) optics. The reduction in beam size when using the quadruplet optics is clearly visible.}
\label{figbeamsize}
\end{center}
\end{figure}

The overfocusing of tail particles is solved by splitting the strong final doublet into a quadruplet. Figure~\ref{figbeamsize} shows the disrupted beam size for the low energy tail particles at the magnetic septum location. This septum is located almost 50m away from 
the IP. These plots were obtained by tracking the disrupted beam along the extraction 
line using an NLC version of DIMAD~\cite{dimad} which performs tracking calculations 
correct to all orders in the energy deviation~$\delta$. This ensures the correct analytic 
treatment of the low energy tail particles. The reduction in disrupted beam size with quadruplet optics can be expected to reduce the losses along the extraction line.

\begin{figure}[h]
\begin{center}
\begin{minipage}{0.48\textwidth}
\includegraphics[width=7.7cm]{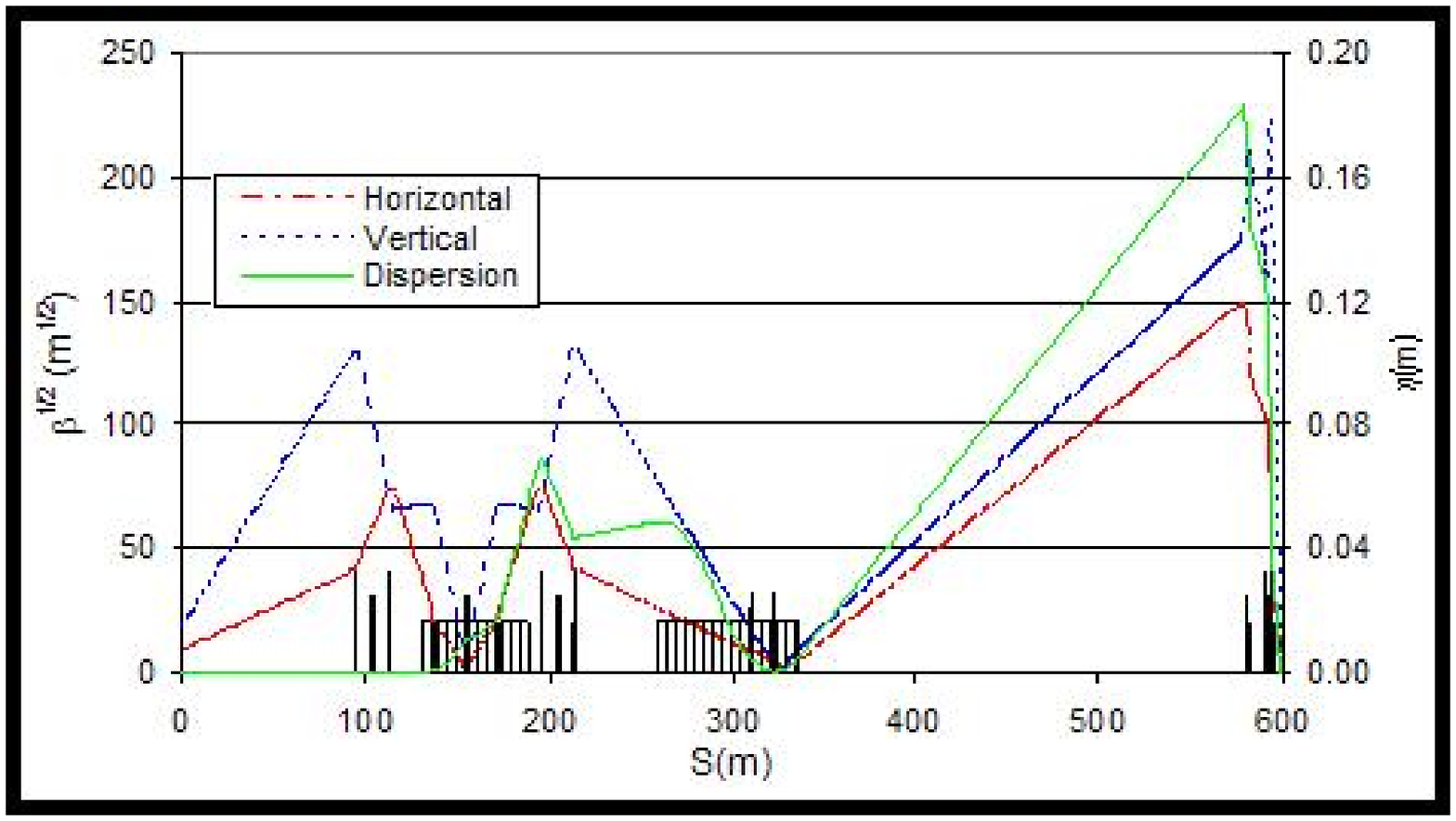}
\end{minipage}
\hfill
\begin{minipage}{0.48\textwidth}
\includegraphics[width=7.7cm]{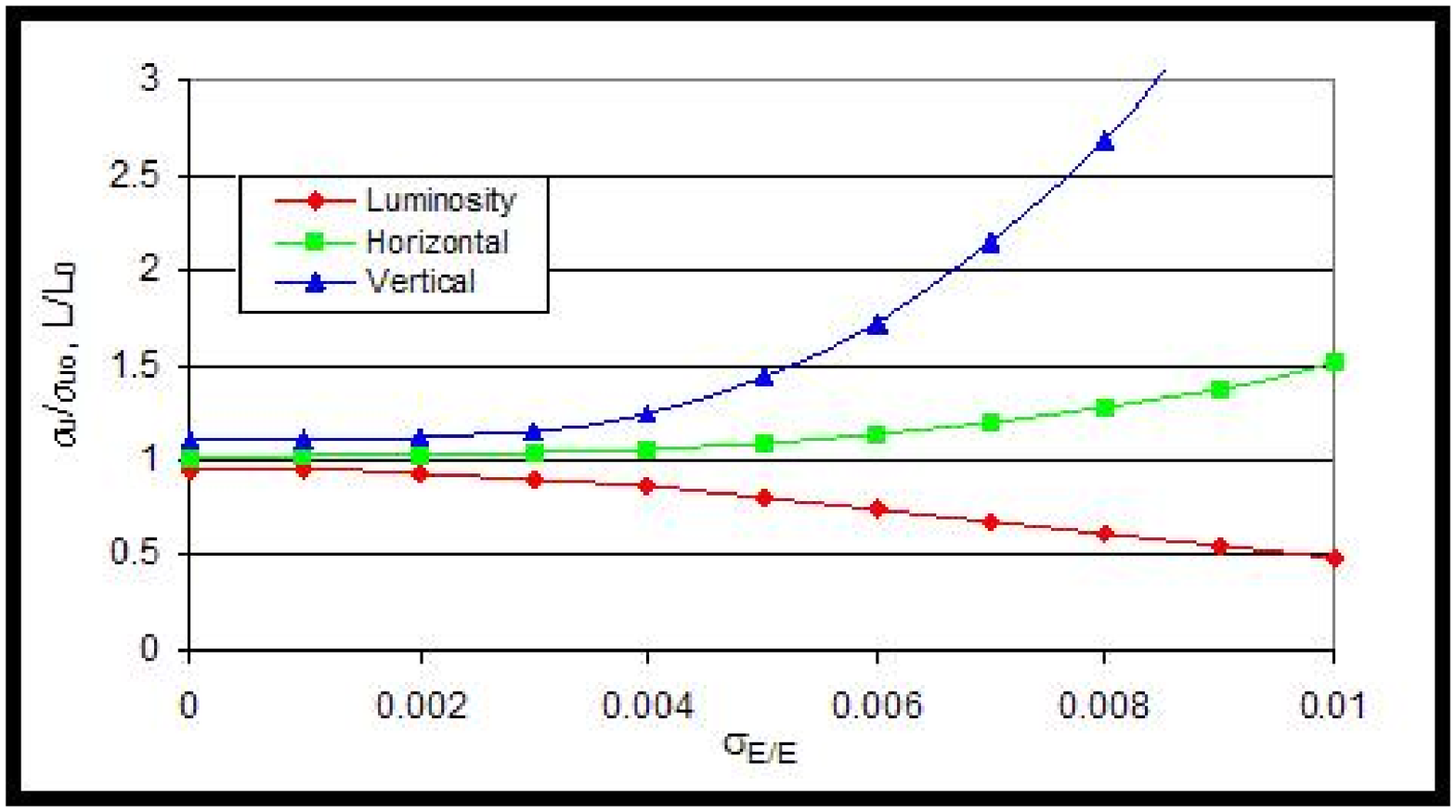}
\end{minipage}
\vspace*{0.5cm}
\caption{\it The left plot shows the TESLA final focus optics for L$^*$=4.1m with a final quadruplet, with the corresponding IP bandwidths shown in the right plot.}
\label{figoptlum}
\end{center}
\end{figure}

The quadruplet optics used to reduce particle losses in the extraction line must satisfy the 
requirements for the incoming beam and must have good chromatic properties. The final focus system with local chromaticity correction~\cite{payet} has been modified to include a final quadruplet and the resulting lattice has been optimised to second order to achieve good chromatic bandwidth. The dipole locations and beam optics have been optimised to keep the horizontal emittance growth due to synchrotron radiation below~$2.5\times 10^{-14}\,\mathrm{m.rad}$. The optical functions of the final focus system and the beam sizes and luminosity as a function of energy spread are shown in the left and right hand plots of figure~\ref{figoptlum}, respectively. The results with the quadruplet are comparable with those obtained using a doublet~\cite{payet}. 

While this scheme resolves the problems found in the extraction of the beamstrahlung photon flux and spent beam, other requirements of the head-on geometry, such as the need for strong electrostatic separators remain. It will also be required to properly mask the synchrotron radiation generated by the off-axis beam in the outgoing quadrupoles to minimise backshining into the detector. The outgoing beams will moreover have offsets in the beam position monitors which may increase the complexity of the IP feedback. A strong crab-crossing correction will moreover be required, something which is not needed in the head-on scheme.

\section{The small horizontal crossing angle scheme}
\label{hori}

The second proposed IR geometry is an adaptation of a scheme studied for CLIC~\cite{napoly}. It has a small $\sim$2mrad horizontal crossing angle and uses two different kinds of quadrupoles for the final doublet: a large bore superconducting r=24mm magnet for the last defocusing quadrupole (QD) and a conventional r=7mm magnet for the next to last focussing element (QF). In this way, the outgoing beam goes through QD horizontally off-axis by about 1mm, which further deflects it away from the incoming beam. The optics for the incoming beam is similar to that described in~\cite{payet}. In the fitted solution, the transport matrix element R$_{22} \simeq 3$ between the interaction point and the exit of QD, resulting in a total angle of $\sim$6mrad between incoming and outgoing beam lines. QD has a length of 1m and a 1.5m drift space is kept between QD and QF. In this way both the outgoing beamstrahlung cone and disrupted charged beam are far enough away from the incoming beam in QF ($\leq $6mm beyond the vacuum chamber at its entrance) so that they can be safely steered in between the pole tips on that side of the magnet. 

\begin{figure}[h]
\begin{center}
\vspace*{-0.6cm}
\includegraphics[width=10cm]{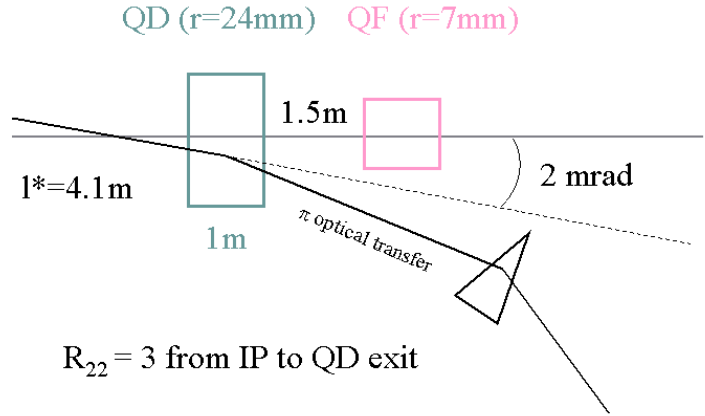}
\vspace*{-0.6cm}
\caption{\it Layout of the IR with a 2mrad horizontal crossing angle.}
\label{figxscheme11}
\end{center}
\end{figure}

\begin{figure}[h!]
\begin{center}
\includegraphics[width=9cm]{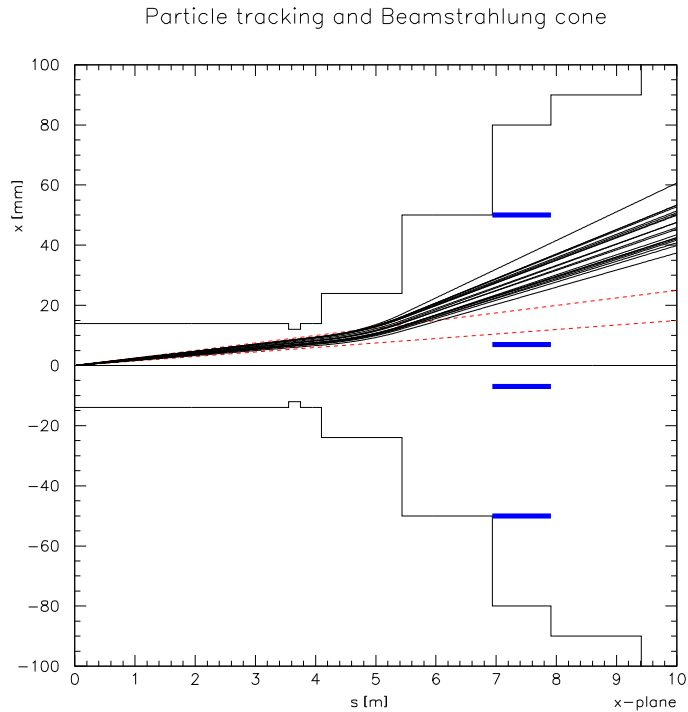}
\caption{\it Schematic of the apertures of the last elements showing the extraction of the beamstrahlung photon flux and charged beam.}
\label{figxscheme12}
\end{center}
\end{figure}

The set-up is sketched in figure~\ref{figxscheme11} together with a schematic of the relevant apertures for both in and outgoing beams up to 10m from the IP (see figure~\ref{figxscheme12}). 


\begin{figure}[h]
\begin{center}
\begin{minipage}{0.48\textwidth}
\includegraphics[width=7.7cm]{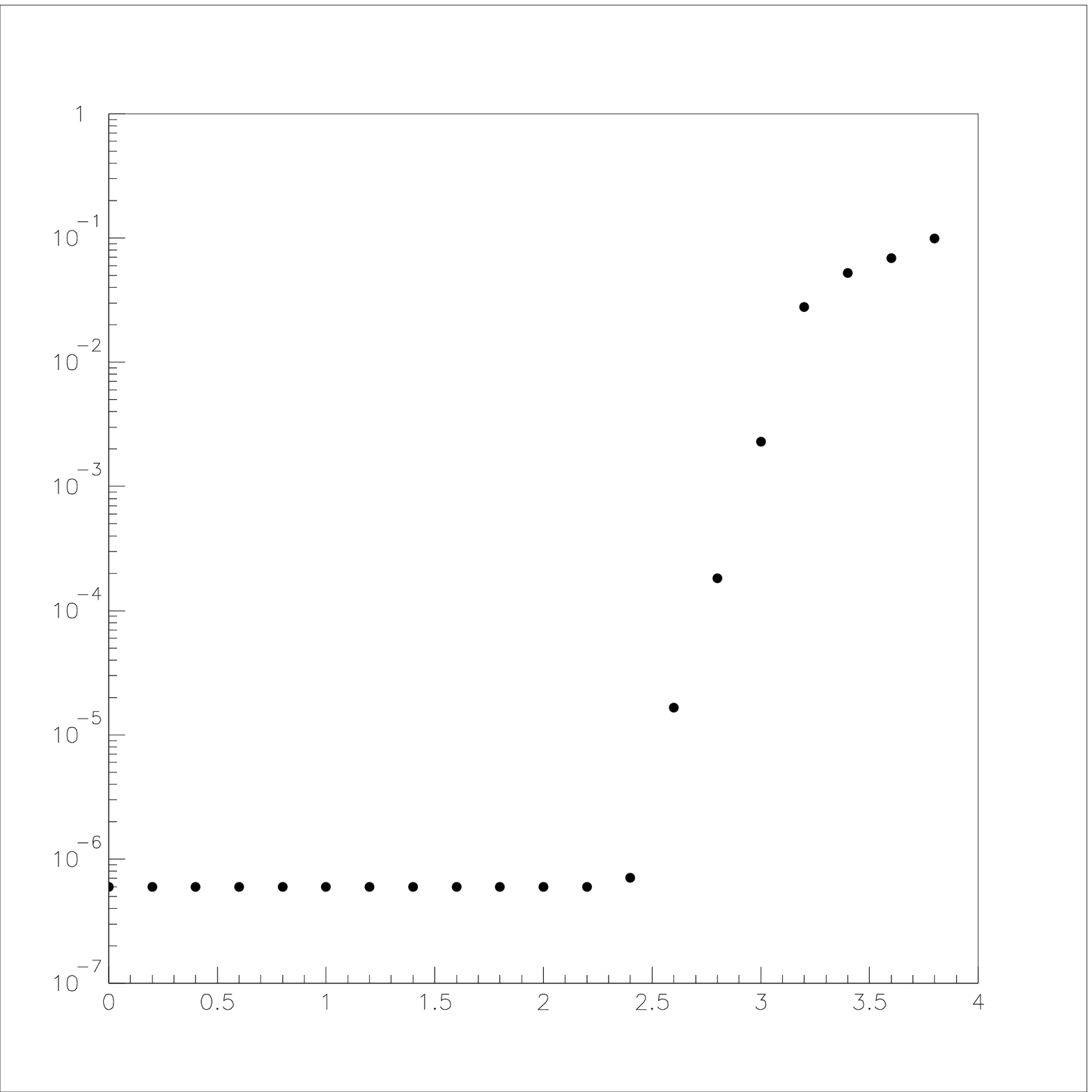}
\end{minipage}
\hfill
\begin{minipage}{0.48\textwidth}
\includegraphics[width=7.7cm]{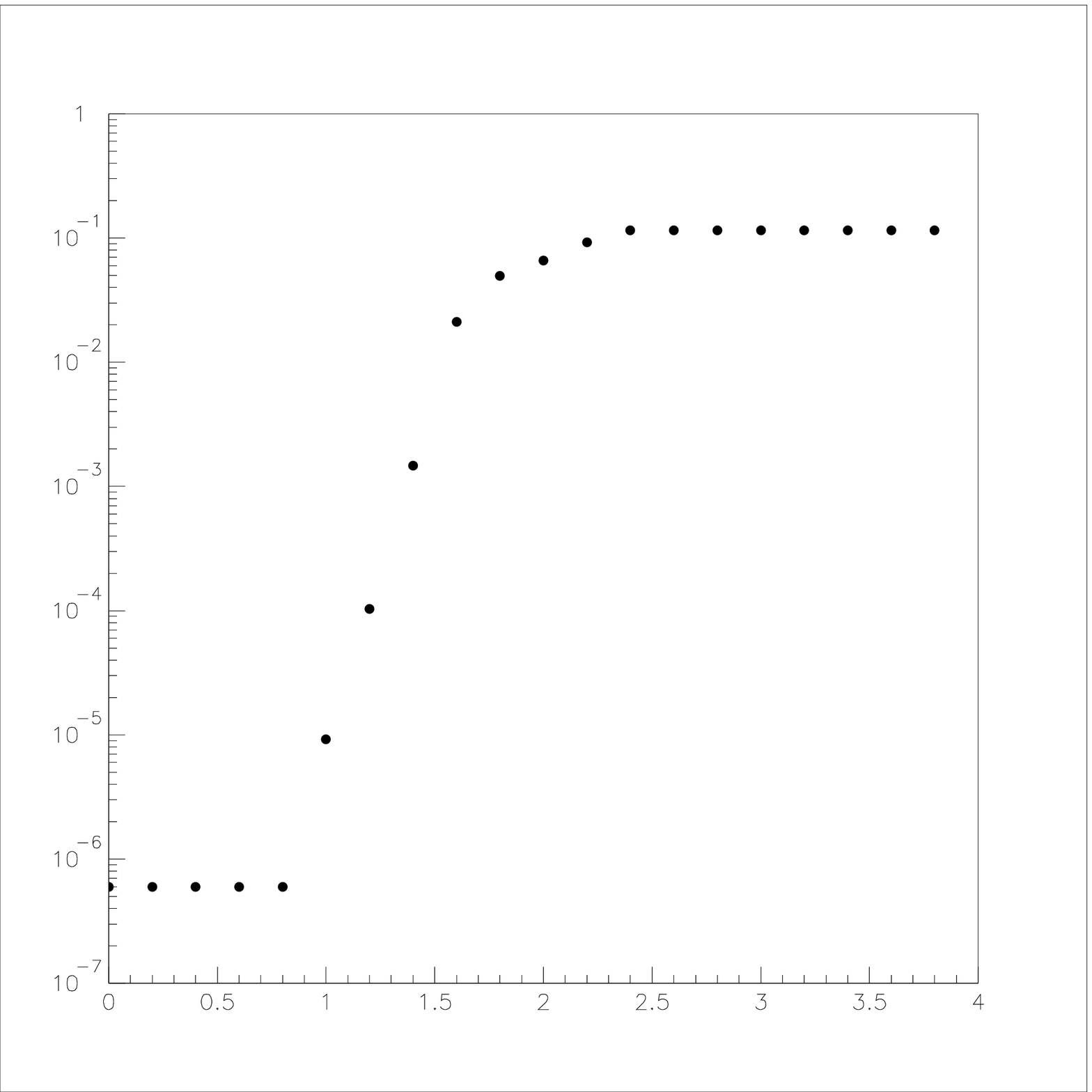}
\end{minipage}
\caption{\it Fraction of outgoing beam power deposited in QD (left) 
and in the entire final doublet QD+QF (right) as a function of total horizontal 
angle in mrad.}
\label{figxscheme2}
\end{center}
\end{figure}

The envelope for the beamstrahlung cone is represented by the dashed lines and corresponds to a $\pm$0.5mrad horizontal angular spread around the 2mrad crossing angle, which is enough to contain most of the emitted power in realistic beam conditions~\cite{napolyschulte}. The tracing
of a representative set of particles from the low-energy tail of the disrupted outgoing beam is 
also depicted in the schematic to illustrate the clearance at the exit of QD. An initial estimate of the fraction of outgoing beam power deposited in QD is shown in figure~\ref{figxscheme2} where it is also compared with the same fraction for the head-on scheme (in the latter case in the entire doublet). Although this study was limited by statistics (the disrupted beam was represented by only 640000 macroparticles) it can be seen that less than 5 $\times 10^{-7}$ of the beam, corresponding to $\simeq$5W at nominal intensity, is deposited in either scheme after passing through the magnetic element(s) common with the incoming beam. This exceeds the 3 W/m limit required to keep the cooling of the superconducting magnet reasonable. Moreover, for the same safety margin to be assured in the crossing angle as in the head-on scheme when taking into account realistic beam conditions, it can be seen from the plots that the crossing angle would have to be limited to $\sim$1.6mrad. A more comprehensive study with more statistics and with a suitable optimisation of both the apertures and lengths is needed to refine these numbers and determine the optimal magnitude for the crossing angle and feasibility of the scheme.

\begin{figure}[h]
\begin{center}
\includegraphics[width=7.7cm]{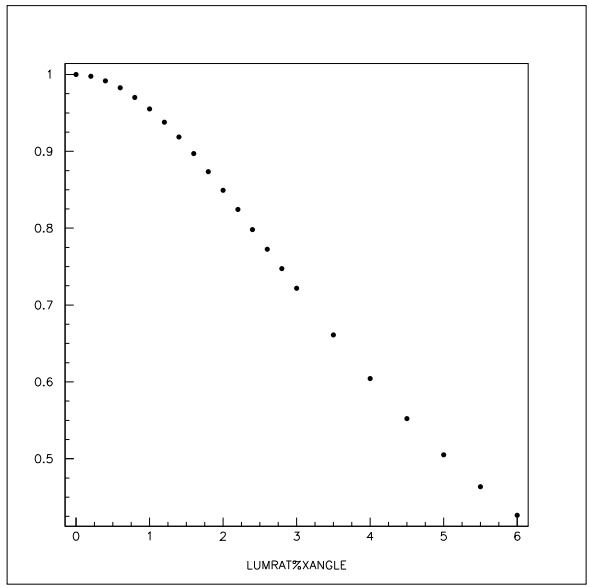}
\caption{\it Luminosity reduction factor as a function of horizontal crossing angle in mrad, obtained with the Guinea-pig simulation for nominal TESLA parameters. For 2mrad, the loss is $\sim 15\%$. This is slightly larger than what can be estimated assuming rigid beams (12\%).}
\label{figxscheme3}
\end{center}
\end{figure}

The rationale for an IR geometry with such a small horizontal crossing angle is as follows. There is no need to develop and operate a very compact high gradient final doublet of quadrupoles (either superconducting of with permanent magnets) as is required for large crossing angles (e.g. 20mrad). Strong electrostatic separators, required for the head-on geometry, are not needed. Only the last quadrupole, QD is common to both beams, instead of the entire QD+QF doublet as in the head-on geometry. This should give a bit more freedom both in the design of 
the optics and operationally. Detrimental effects on the physics program (e.g. reduced hermeticity in the forward region, complications from the solenoid and beam axes not being aligned) are negligible. With a horizontal crossing angle of $\sim$2mrad and for nominal TESLA parameters~\cite{teslatdr}, only $\sim$15\% of the luminosity is lost without using crab-crossing (see Figure~\ref{figxscheme3}, obtained using the Guinea-pig simulation~\cite{napolyschulte}), compared to a factor of about 5 for a 20mrad crossing angle. Correction of this 15\% loss may be possible without dedicated cavities, by exploiting the angular dispersion required at the collision point in the design of the final focus optics~\cite{payet} to enable local chromaticity correction. Diagnostics of the spent beam should be easier than in the head-on scheme, although it remains to be checked that a polarimeter and energy spectrometer can both be designed with suitable performances in the outgoing beam line.

\section{Conclusion}
\label{conc}

Two options have been suggested to save the advantages of the head-on collision scheme proposed in the TESLA design. In the first, a small vertical crossing angle ($\sim$0.3mrad) at the IP can alleviate the problem of beamstrahlung heating of the septum blade. To reduce the low energy tail particle losses the strong final doublet can be replaced by a quadruplet. A final focus system with good chromatic properties can been designed with such a quadruplet. However, as for the head-on scheme, R\&D on electrostatic separators will be needed, especially for the upgrade to 1 TeV. Moreover a strong crab-crossing correction is required to maintain the luminosity in this scheme. The second option uses a small ($\sim$2mrad) horizontal crossing angle. This scheme is attractive as it does not need electrostatic separators and requires only a very modest crab-crossing correction, which moreover may be achieved without special cavities, by exploiting finite dispersion at the IP. Many details of both designs still need to be worked out, including optimising the magnitude of the crossing angle in the second scheme, confirming that power losses in the extraction channel are tolerable and studying whether suitable post-IP diagnostics can be included.

\end{document}